# Widefield Spectroscopic Telescope (WST): coating strategy to achieve high optical throughput

Benoit Sassolas[1], Matthieu Coulon[1], Christophe Michel[1], Laurent Pinard[1], Roland Bacon[2], Corentin Cudennec[2], Philippe Dierickx[2]

[1] Laboratoire Matériaux Avancés-IP2I, CNRS UCB Lyon, 69100 Villeurbanne, France
[2] Université Claude Bernard Lyon 1, CNRS, Centre de Recherche Astrophysique de Lyon, UMR5574

## ABSTRACT

The Wide-field Spectroscopic Telescope (WST) is a 12-m class facility designed for simultaneous wide-field multi-object and integral-field spectroscopy across 370–1600 nm, targeting a throughput above 83% its wide-field focus and 76% at its integral field one. Its 13 mirrors and 5 lenses require optimized coatings balancing feasibility, operational needs, and long-term durability. Large mirrors (>2 m) may use enhanced metallic coatings, such as the protected-silver solution from the Vera C. Rubin Observatory, achieving over 90% reflectivity in UV and above 99% in NIR. Smaller mirrors would use $Nb_2O_5/SiO_2$ dielectric stacks, offering ~99% reflectivity, tunable spectral response, and excellent stability. Coating the large lenses (up to 1.6 m) is challenging due to the ultra-broadband range; traditional multilayer antireflective coatings show limitations from manufacturing and incidence-angle effects. Graded-index AR coatings are being explored for superior performance across broad wavelengths and angles.

## 1. INTRODUCTION

The Wide-field Spectroscopic Telescope (WST) is a 12-m class facility designed to perform simultaneous wide-field multi-object and integral-field spectroscopy [1]. The wide-field focal surface (MOS) is reached after 2 reflecting and 6 refracting surfaces, while the integral-field (IFS) one is reached after 11 additional reflecting and 4 refracting surfaces (see Figure 1). The target efficiency is 83 % over 0.37-1.60 µm at the wide-field focus, and 76% over the range 0.37-0.93 µm at the integral field one. Other requirements are listed in Table 1.

To meet this ambitious goal despite the large number of optics, particular attention is devoted to defining an optimal coating strategy that accounts for feasibility, operational constraints and long-term sustainability. Performances in the near ultraviolet is a design driver for optical coatings since common materials exhibit higher optical absorption. This effort is a twofold activity

- Increase reflectivity of the mirrors above 98%
- Reduce the Fresnel loss on the lens surfaces below 1%

This paper presents the coating strategy worked out and the preliminary developments to validate the proposed coatings solutions.

Table 1: Specification of the average efficiency ($Y_{ave}$) and minimal efficiency ($Y_{min}$) for each observing path of WST

| Path | Spectral range [nm] | $Y_{ave}$ | $Y_{min}$ |
|---|---|---|---|
| MOS | 370 − 1600 | 83 % | 64% |
| IFS | 370 − 930 | 76 % | 66 % |

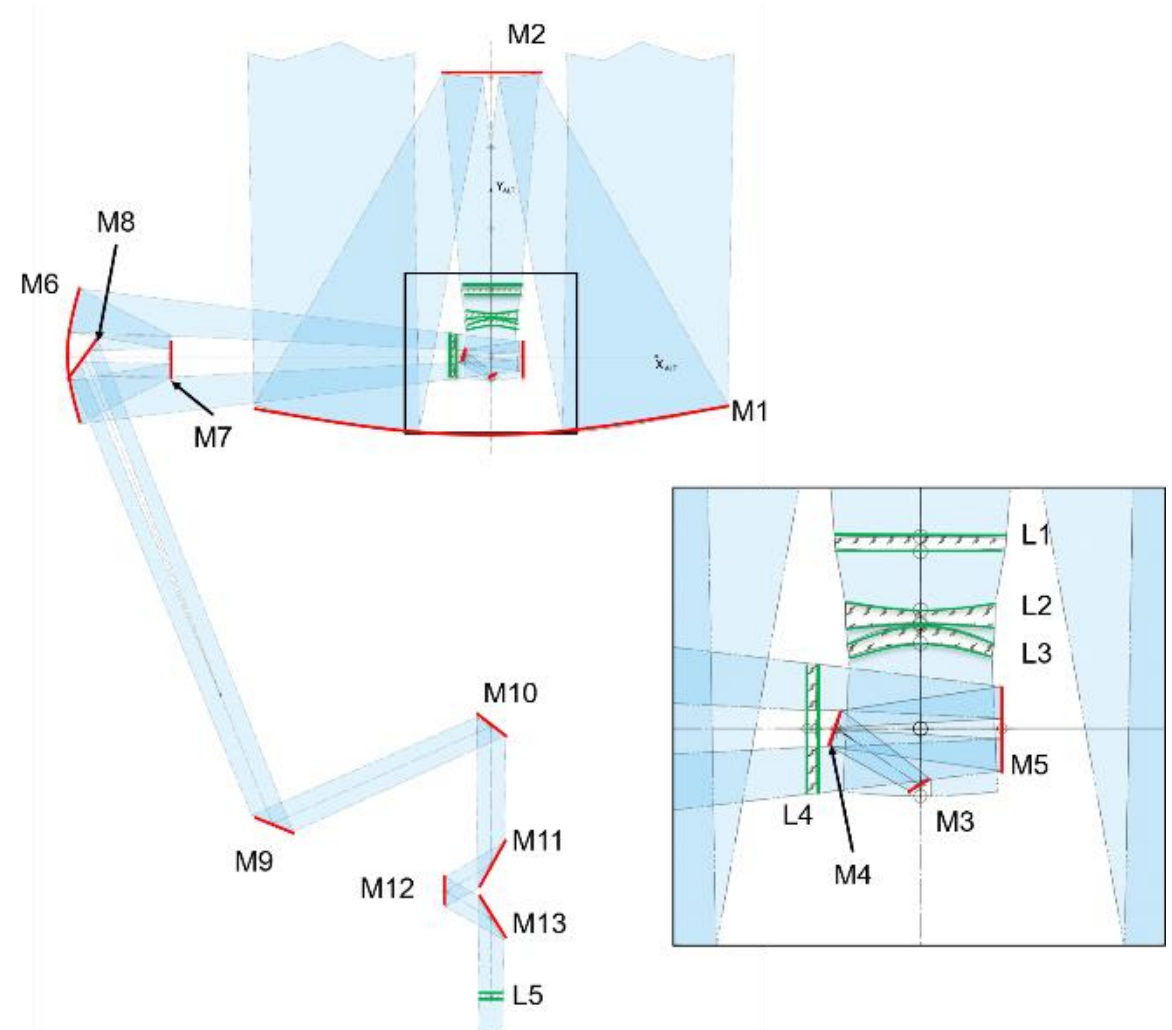

Figure 1: Optical layout of WST with reflective surfaces in red and transmissive surfaces in green. Credit: WST/P. Dierickx

# 2. METHODS

## 2.1 Mirrors

Classical metallic coatings such as protected aluminum or protected silver exhibit either low average reflectance or sharp decrease in the UV. Their limited performance would have a strong impact in WST since there may be up 13 mirrors depending on the path (i.e. MOS or IFS). Hence, other coating solutions have to be defined.

The mirrors M2 and M6 are highly critical because of their diameter of 2.5 m and 3.5 m, respectively. For risk mitigation, the state-of-the-art enhanced metallic coatings already used in the current large telescopes are considered. Outstanding performance for the M1/M3 mirror of the Vera C. Rubin Observatory (8.4 m in diameter), was demonstrated with an enhanced silver coating based on chromel and silicon nitride [2]. Reflectance above 90% in the UV and more than 99% in the NIR has been achieved (see Figure 2).

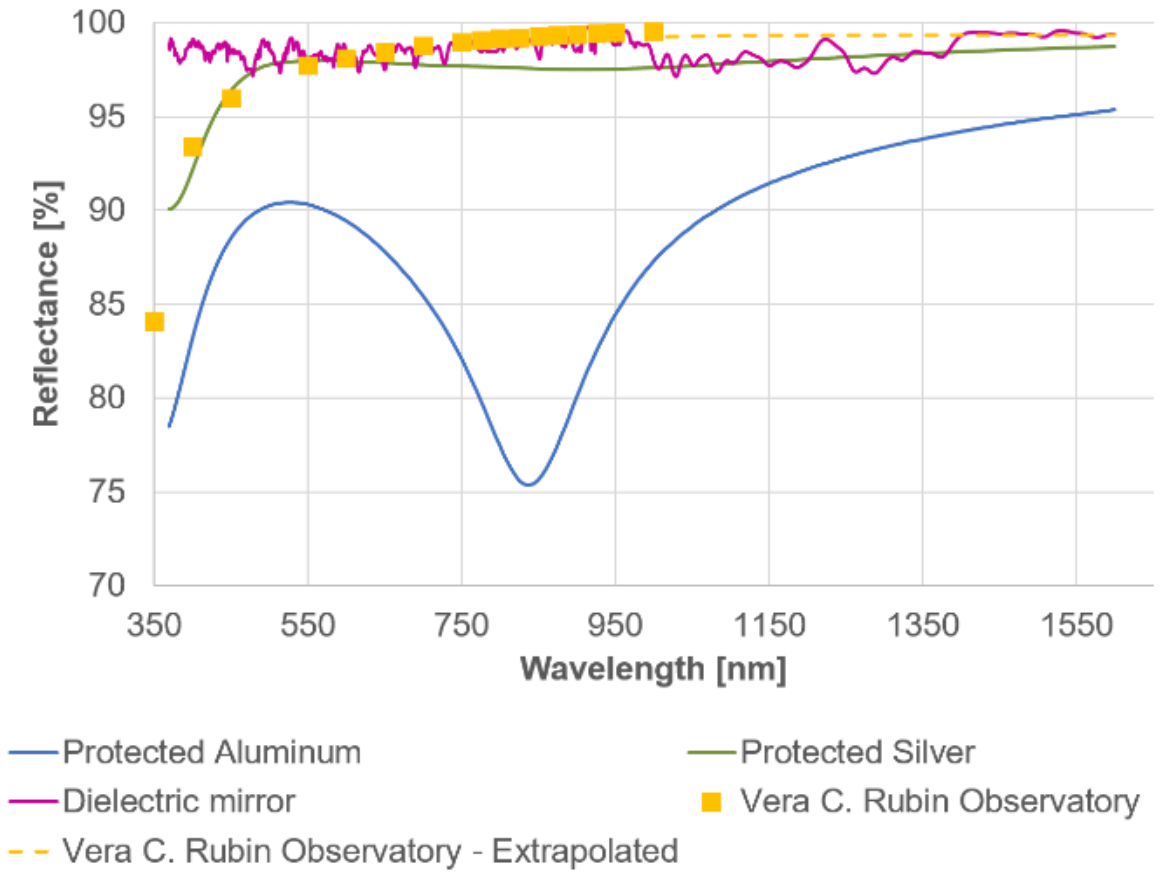

Figure 2 : Reflectance of various reflective coatings. Dielectric coating is based on $Nb_2O_5/SiO_2$ multilayer. Vera C. Rubin Observatory from [2]

All-dielectric multilayers are proposed for other mirrors. This technical solution based on interferential effects in thin-films stack enables to tune the reflectivity according to the spectral range and the angle of incidence [3]. Low absorption is possible thanks to lossless materials and high reflectivity can be reached quite easily: the more the layers, the more the reflectance. However, slight absorption in the UV is vey likely and special attention is required to the materials and the design. For WST, multilayers based on $Nb_2O_5$ and $SiO_2$ appear to be an interesting solution. The large contrast between the refractive index of $Nb_2O_5$ (2.25) and $SiO_2$ (1.45) offers a high reflectance at the interfaces. Moreover, absorption below 0.5% in the UV seems to be a realistic target even for mirrors with the broadest reflection zone. Mirrors made up $Nb_2O_5$ and $SiO_2$ with average reflectance higher than 98.5% seem then a reasonable baseline.

### 2.2 Lenses

A simple sensitivity analysis highlighted that the total throughput of the telescope is actually limited by the transmission of the lenses. In this case, the reflectivity has to be reduced as low as possible. Classical anti-reflective coatings (AR) using multilayer have good performance but they are intrinsically limited by the width of the spectral range [4] or polarization effects at oblique incidence [3]. Graded index layer (GRIN) coatings have quite recently shown promising results in the visible range. Low residual reflectance over a broad spectral range and excellent omnidirectionality have been demonstrated with a GRIN coating called “grass-like” $Al_2O_3$ [5,6]. Such an AR coating with an average transmittance close to 99.5% would provide significant improvements compared to multilayer AR solutions (see Figure 3).

This structure was initially developed for solar cells applications and further developments are required to validate its relevance to large optical lenses, especially regarding the durability. Dedicated tests are on-going at LMA and preliminary results are reported in section 3.2.

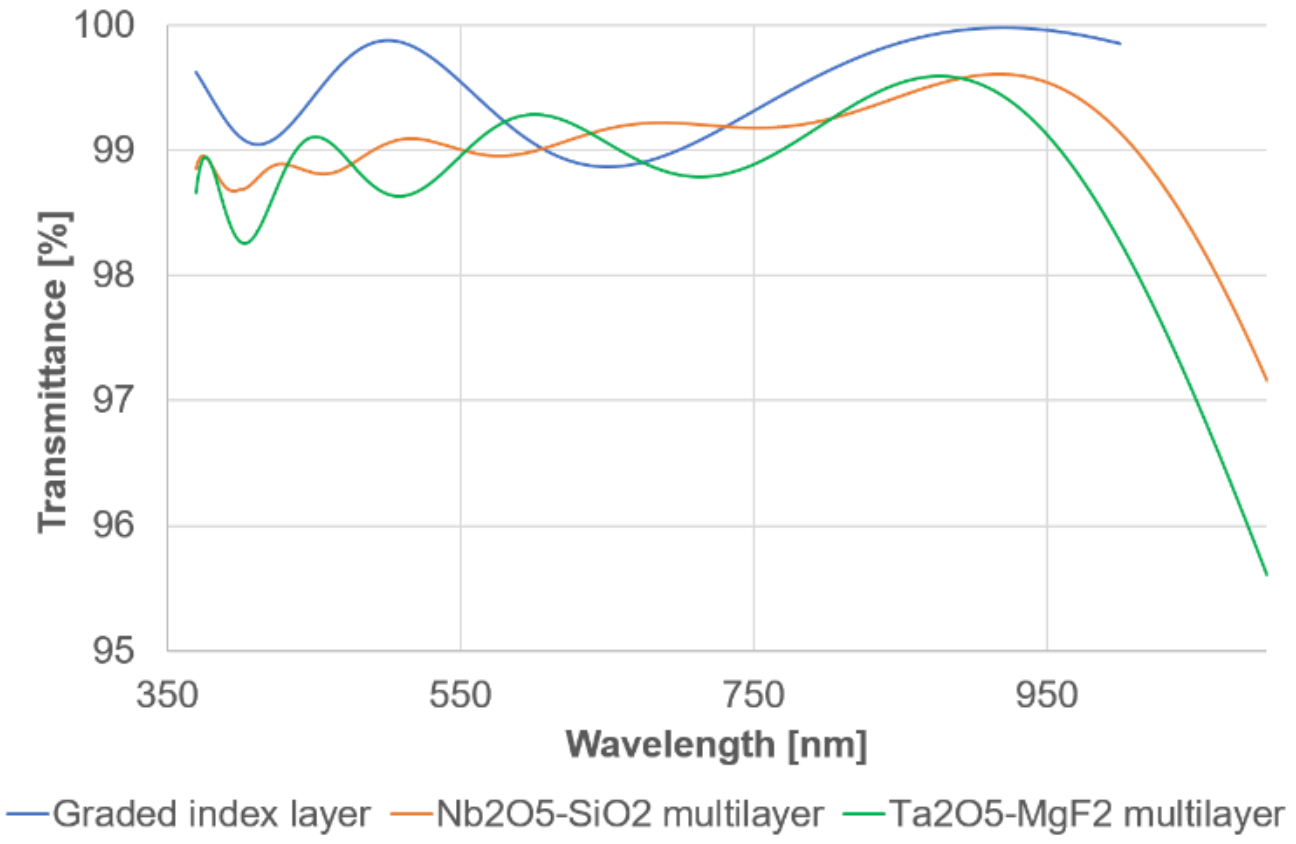

Figure 3: Reflectance of different AR coatings. Graded index layer inspired by “grass-like” $Al_2O_3$ layer, and multilayer AR based on $Nb_2O_5/SiO_2$ or $Ta_2O_5/MgF_2$ layers.

## 3. RESULTS

### 3.1 Optical throughput

The optical throughput has been computed by considering enhanced silver coatings from [2] for M2 and M6 and all-dielectric coating for other mirrors. Two options are then considered for the AR coatings, $Ta_2O_5$-$MgF_2$ AR coating and grass-like $Al_2O_3$.

Tableau 2 : Expected throughput of WST for different coating solutions

| | Path | $Y_{ave}$ | $Y_{min}$ |
|---|---|---|---|
| $Ta_2O_5$-$MgF_2$ AR | MOS | 88 % | 77 % |
| | IFS | 71 % | 61 % |
| Grass-like $Al_2O_3$ | MOS | 91 % | 83 % |
| | IFS | 79 % | 71 % |

As expected, AR coating based on grass-like $Al_2O_3$ improves significantly the total throughput, especially for the IFS path with an increase of 10% on the average and minimum values. The specifications are then satisfied with some comfortable margins.

### 3.2 Preliminary tests on grass-like $Al_2O_3$

Grass-like alumina is a "high risk – high gain" solution. As reported in [5,6], the structure is obtained thanks to a post-deposition treatment of an alumina layer with deionized water. But it is useful to remind that this technology requires a dedicated development plan in order to be validated for astronomical applications. The very preliminary results are presented in this section.

A ∅1'' substrate in fused silica was coated with an alumina layer using ion-beam sputtering technique. A target of pure aluminum was sputtered with a 1 keV $Ar^+$ ion beam under vacuum with $10^{-4}$ mbar oxygen partial pressure. The thickness of the as-deposited layer is about 50 nm. The transmittance was measured after several post-deposition treatments with deionized water (see Figure 4). The minimum transmission has increased from 86.5% to 94.5% in only 3 steps. By taking into account the contribution of the uncoated backside, we can estimate the average transmittance of the AR coating close to 98.8%. Further research is necessary to attain the 99.5% benchmark, though the current result constitutes a promising initial outcome.

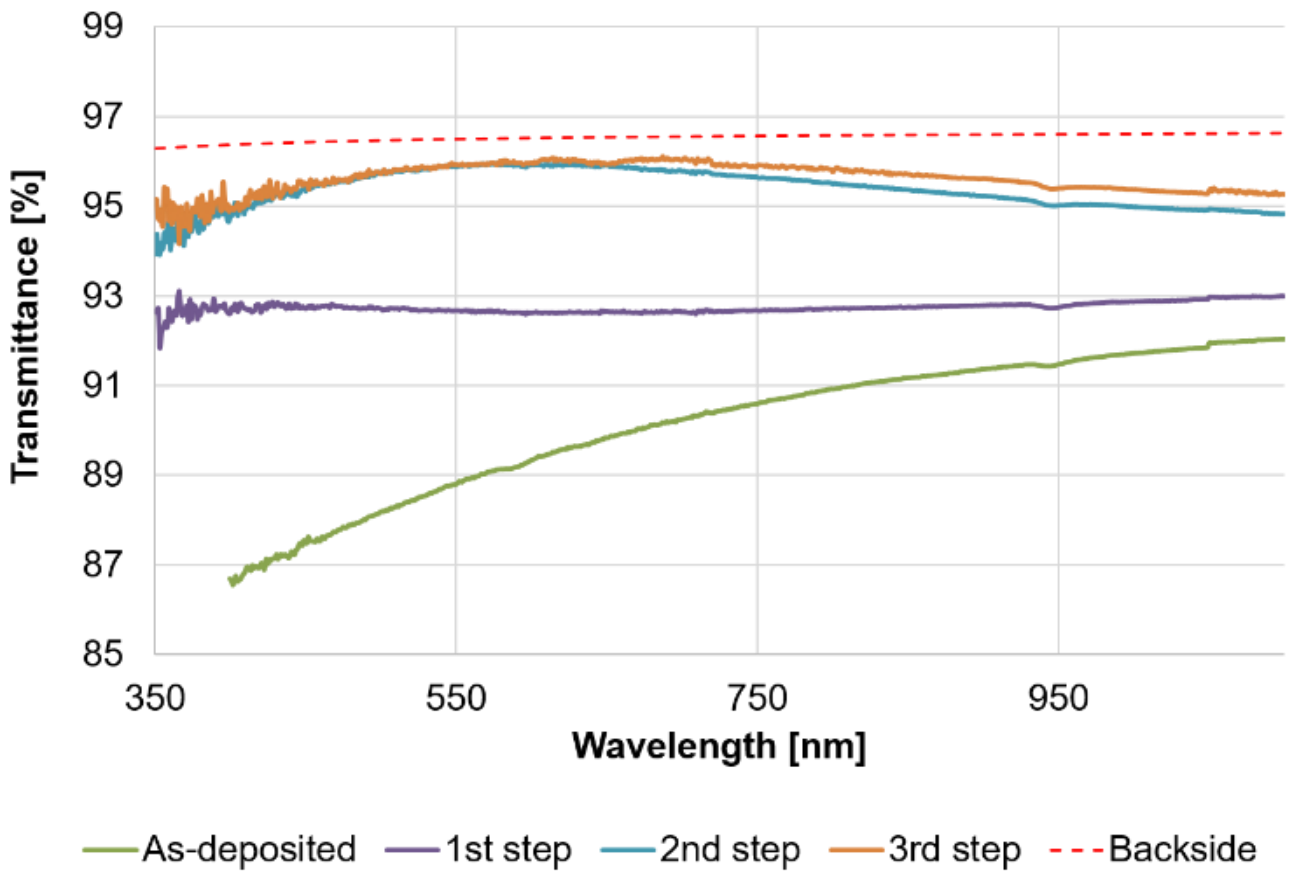

Figure 4: Transmittance of grass-like alumina layer at different steps of the post-deposition process. "Backside" is the transmittance of the uncoated backside of the sample.

## 4. CONCLUSION

WST is an ambitious project that aims performing simultaneous wide-field multi-object and integral-field spectroscopy. The large number of optics in the optical train is a critical point because of the fraction light lost at each surface. A special effort is then focused in defining the best coating strategy in order to maximize the optical throughput. Encouraging results are achieved by using enhanced silver coatings, all-dielectric multilayers as reflective coatings and grass-like alumina as anti-reflection coating.

## ACKNOWLEDGEMENTS

This project has received funding from the European Union Horizon Europe Research and Innovation Action under grant agreement no. 101183153 -WST.